\documentclass[prx,onecolumn,notitlepage,showpacs,longbibliography,superscriptaddress]{revtex4-1}
\usepackage{amsmath,amssymb,bm,mathrsfs,graphicx, braket,amsthm,enumerate,times}
\usepackage[colorlinks=true,citecolor=blue,linkcolor=blue]{hyperref}

\newcommand{\ed}[1]{#1}
\newcommand{\edd}[1]{#1}

\begin{document}
\title{Proof of the absence of long-range temporal orders in Gibbs states}
\author{Haruki Watanabe}
\email{haruki.watanabe@ap.t.u-tokyo.ac.jp}
\affiliation{Department of Applied Physics, University of Tokyo, Tokyo 113-8656, Japan}

\author{Masaki Oshikawa}
\email{oshikawa@issp.u-tokyo.ac.jp}
\affiliation{Institute for Solid State Physics, University of Tokyo, Kashiwa 277-8581, Japan} 

\author{Tohru Koma}
\email{tohru.koma@gakushuin.ac.jp}
\affiliation{Department of Physics, Gakushuin University, Mejiro, Toshima-ku, Tokyo 171-8588, Japan} 

\begin{abstract}
\edd{We address the question whether time translation symmetry can be spontaneously broken in a quantum many-body system. One way of detecting such a symmetry breaking is to examine the time-dependence of a correlation function. }
If the large-distance behavior of the correlation function exhibits a nontrivial time-dependence in the thermodynamic limit, the system would develop a temporal long-range order, realizing a time crystal. In an earlier publication, we sketched a proof for the absence of such time dependence in the thermal equilibrium described by the Gibbs state [H. Watanabe and M. Oshikawa, Phys. Rev. Lett. 114, 251603 (2015)]. Here we present a complete proof and extend the argument to a more general class of stationary states than the Gibbs states.
\end{abstract}

\maketitle

\section{Introduction}
Time crystals are a newly proposed state of matter that spontaneously breaks the time translation symmetry.
The idea of time crystals in the case of \edd{the continuous time translation symmetry} was first proposed by Wilczek in 2012~\cite{Wilczek}, although the validity of the concrete model in this original proposal was soon questioned in Ref.~\cite{BrunoComment1}.  Then a no-go theorem for a wider but still restricted class of models was presented in Ref.~\cite{Bruno}.  In a more general setting, the absence of time crystalline orders in the ground state or in the Gibbs state was proven in Ref.~\cite{WO} without specifying the Hamiltonian but assuming only its locality.  These developments triggered further investigation of so-called Floquet time crystals or discrete time crystals in nonequilibrium setting~\cite{Sacha,Khemani,Else,Yao,TC1,TC2} that break a discrete time translation symmetry into its subgroup.  See Refs.~\cite{SachaReview,ElseReview,SondhiReview} for recent reviews on this topic.

The argument for the no-go theorem at finite temperatures in Ref.~\cite{WO} was based on the Lieb-Robinson bound~\cite{LRB,HastingsReview}, which was used to constrain finite time behavior of the correlation function. However, in Ref.~\cite{WO}, Fourier transformation of the correlation function was performed with respect to an infinitely long time, out of the validity of the constraint. This issue was recently pointed out by Ref.~\cite{SondhiReview}.  In this work, we present a complete version of the proof without such an issue. Furthermore, we \edd{examine} the conditions on the density operator to which our argument can be straightforwardly extended. Clarifying these subtleties and settling down the limitations on what the Gibbs state and similar type of stationary states can do should in turn accelerate our exploration of new states that exhibit nontrivial temporal orders.

\section{The theorem and its proof}
\subsection{Setup and statement}
Let us consider a static Hamiltonian $\hat{H}$ defined on a \edd{$d$-dimensional lattice $\Lambda$ that is a finite subset of $\mathbb{Z}^d$.  }
We assume that the Hamiltonian $\hat{H}$ is written as a sum of local bounded Hamiltonians:\edd{
\begin{equation}
\hat{H}=\sum_{\vec{x}\in\Lambda}\hat{h}_{\vec{x}}.
\end{equation}
More precisely, we assume that the support of the local Hamiltonian $\hat{h}_{\vec{x}}$ is limited to a finite range $R_h$ from $\vec{x}\in\Lambda$ and that the operator norm~\footnote{The operator norm of an operator $\hat{O}$ is defined as ${\Vert}\hat{O}{\Vert}:=\text{sup}_{|\psi\rangle, {\Vert}{|\psi\rangle}|{\Vert}>0}{\Vert}\hat{O}{|\psi\rangle}{\Vert}/{\Vert}{|\psi\rangle}{\Vert}$.} of $\hat{h}_{\vec{x}}$ is bounded by a constant $N_h$. Both $R_h$ and $N_h$ are independent of the position $\vec{x}\in\Lambda$ or the system size $|\Lambda|$. This setting includes a wide variety of quantum spin systems, fermion systems, and ``hard-core" boson systems~\footnote{For the assumed boundedness, the maximum number of bosons that can occupy a single site must be a finite number independent of the system size}.}

\edd{Similarly, we consider observables (not necessarily Hermitian) $\hat{A}$ and $\hat{B}$ written as a sum of local observables:
\begin{align}
\hat{A}:=\frac{1}{|\Lambda|}\sum_{\vec{x}\in\Lambda} \hat{a}_{\vec{x}},\quad\hat{B}:=\frac{1}{|\Lambda|}\sum_{\vec{x}\in\Lambda} \hat{b}_{\vec{x}}.
\end{align}
The support of $\hat{a}_{\vec{x}}$ and $\hat{b}_{\vec{x}}$ ($\vec{x}\in\Lambda$) are within a finite range $R_a$, $R_b$ from $\vec{x}$ and their operator norm is bounded by constants $N_a$, $N_b$, respectively. All of these constants are independent of $\vec{x}$ or $|\Lambda|$.}

We introduce the time evolution of operators for $t\in\mathbb{R}$ by
\begin{equation}
\hat{A}(t):=e^{i\hat{H} t}\hat{A}e^{-i\hat{H} t}.
\end{equation}
Our interest is in the time-dependence of the correlation function
\begin{equation}
\langle\hat{A}(t)\hat{B}\rangle:={\rm Tr}\left(\hat{A}(t)\hat{B}\hat{\rho}\right).\label{fpdef}
\end{equation}
Here $\hat{\rho}$ is the Gibbs state
\begin{equation}
\hat{\rho}:=\frac{1}{Z}e^{-\beta \hat{H}}\label{Gibbs}
\end{equation}
at the inverse temperature $\beta$ and $Z:={\rm Tr}\,e^{-\beta \hat{H}}$ is the partition function.  Our claim is that $\langle\hat{A}(t)\hat{B}\rangle$ is independent of $t$ in the thermodynamic limit $|\Lambda|\rightarrow+\infty$~\cite{WO}, i.e.,
\begin{equation}
\lim_{|\Lambda|\rightarrow\infty}\left|\langle\hat{A}(t)\hat{B}\rangle-\langle\hat{A}\hat{B}\rangle\right|=0.\label{result}
\end{equation}

\subsection{Proof for $\beta>0$}
\label{secbetap}

\subsubsection{Outline}
\label{outline}
To prove Eq.~\eqref{result}, it is sufficient \edd{to treat} the special case $\hat{B}=\hat{A}^\dagger$:
\begin{equation}
\lim_{|\Lambda|\rightarrow\infty}\left|\langle\hat{A}(t)\hat{A}^\dagger\rangle-\langle\hat{A}\hat{A}^\dagger\rangle\right|=0.\label{result2}
\end{equation}
This is because $\langle\hat{A}(t)\hat{B}\rangle$ can be rewritten as
\begin{align}
\langle\hat{A}(t)\hat{B}\rangle=
\frac{1}{2}\Big\langle\big(\hat{A}(t)+\hat{B}^\dagger(t)\big)\big(\hat{A}+\hat{B}^\dagger\big)^\dagger\Big\rangle
+\frac{i}{2}\Big\langle\big(\hat{A}(t)+i\hat{B}(t)^\dagger\big)\big(\hat{A}+i\hat{B}^\dagger\big)^\dagger\Big\rangle
-\frac{1+i}{2}\langle\hat{A}(t)\hat{A}^\dagger\rangle
-\frac{1+i}{2}\langle\hat{B}^\dagger(t)\hat{B}\rangle.
\end{align}
Once Eq.~\eqref{result2} is established, it applies to all four correlation functions in the right-hand side and we obtain Eq.~\eqref{result}.

We denote by $|\Phi_n\rangle$ the \edd{eigenstate} of the Hamiltonian $\hat{H}$ with the eigenvalue $E_n$ ($n\in\mathbb{N}$).  Using the complete system and writing 
\begin{align}
\rho(E_n):=\frac{1}{Z}e^{-\beta E_n},\label{Gibbs2}
\end{align}
we get 
\begin{align}
\langle\hat{A}(t)\hat{A}^\dagger\rangle=\sum_{m,n}|\langle\Phi_m|\hat{A}|\Phi_n\rangle|^2\rho(E_m)\,e^{i(E_m-E_n)t}.
\end{align}
We split the summation over $m$ and $n$ into four intervals of $E_n-E_m$:
\begin{equation}
\begin{cases}
\text{(i):}&2\varepsilon\leq E_n-E_m\leq K,\\
\text{(ii):}&-K\leq E_n-E_m\leq -2\varepsilon,\\
\text{(iii):}&K<|E_n-E_m|,\\
\text{(iv):}&|E_n-E_m|<2\varepsilon,
\end{cases}
\end{equation}
where $\varepsilon$ is a small positive number and $K$ is a large positive number. 
Then the time-dependence of $\langle\hat{A}(t)\hat{A}^\dagger\rangle$ can be bounded as
\begin{align} 
\left|\langle\hat{A}(t)\hat{A}^\dagger\rangle-\langle\hat{A}\hat{A}^\dagger\rangle\right|
\leq{}  &  2\sum_{m,n\,:\,2\varepsilon\leq E_n-E_m\leq K}|\langle \Phi_m|\hat{A}|\Phi_n\rangle|^2\rho(E_m)\notag\\
&+2\sum_{m,n\,:\,-K\leq E_n-E_m\leq -2\varepsilon}|\langle \Phi_m|\hat{A}|\Phi_n\rangle|^2\rho(E_m)\notag\\
&+2\sum_{m,n\,:\,K<|E_n-E_m|}|\langle\Phi_m|\hat{A}|\Phi_n\rangle|^2\rho(E_m)\notag\\
&+\sum_{m,n\,:\,|E_n-E_m|<2\varepsilon}|\langle \Phi_m|\hat{A}|\Phi_n\rangle|^2\rho(E_m)\left|e^{i(E_m-E_n)t}-1\right|.\label{f+expand}
\end{align}
In the following, we derive an upper bound for each term in the right hand side one by one.
The first two terms will be bounded using the Lieb-Robinson bound and the monotonically decreasing nature of the Boltzmann factor \eqref{Gibbs2}.
The third term will be evaluated by making use of the large energy difference.
Finally, the last term is trivially small because of the time-dependent factor with small energy difference.  Plugging these results [Eqs.~\eqref{difE+bound}, \eqref{difE-bound},  \eqref{difElargeK}, and \eqref{smallEdef} below] into the right-hand side of Eq.~\eqref{f+expand}, we get
\begin{align} 
\left|\langle\hat{A}(t)\hat{A}^\dagger\rangle-\langle\hat{A}\hat{A}^\dagger\rangle\right|
&\leq 2\varepsilon+2\varepsilon+\frac{2C}{K^2}+2N_a^2\varepsilon |t|,
\label{combine}
\end{align}
where $C$ is a positive constant independent of the system size. 
Since we can take $\varepsilon$ to be small and $K$ to be large by choosing a sufficiently large system size $|\Lambda|$, we obtain the desired result.

\subsubsection{The range (i): $2\varepsilon\leq E_n-E_m\leq K$}
Let us start with the contribution from the range $2\varepsilon\leq E_n-E_m\leq K$. To this end, we introduce a cutoff function $\eta^{+}\in C_0^\infty(\mathbb{R})$ (i.e., an infinitely differentiable function with a compact support ) that satisfies the following conditions:~\footnote{An example of $\eta^{+}(\omega)$ for the range $\varepsilon\leq\omega\leq2\varepsilon$ and $K\leq\omega\leq K+\varepsilon$ can be constructed using $m(x):=\int_{-1}^{x}dy\,e^{-\frac{1}{1-y^2}}$ ($-1\leq x\leq+1$). For example, one can set $\eta^{+}(\omega)=m(\frac{2\omega-3\varepsilon}{\varepsilon})/m(+1)$ for $\varepsilon\leq\omega\leq2\varepsilon$.}
\begin{align}
\begin{cases}
\eta^{+}(\omega)=1 &  (2\varepsilon\leq\omega\leq K),\\
\eta^{+}(\omega)=0 &  (\omega\leq\varepsilon\text{ or }K+\varepsilon\leq \omega),\\
0\leq\eta^{+}(\omega)\leq1&  (\text{otherwise}).
\end{cases}\label{etap}
\end{align}
The Fourier transform of $\eta^+(\omega)$ is given by 
\begin{equation}
\tilde{\eta}^+(t):=\frac{1}{2\pi}\int_{-\infty}^{+\infty}d\omega\, e^{i\omega t}\eta^+(\omega),
\end{equation}
which decays faster than any power of $t$. This can be shown by performing an integration by parts repeatedly:
\begin{equation}
\tilde{\eta}^+(t)=\frac{1}{2\pi}\left(\frac{i}{t}\right)^\ell\int_{-\infty}^{+\infty}d\omega\,
e^{i\omega t}\frac{\partial^\ell}{\partial \omega^\ell}\eta^+(\omega)\quad \text{ for $t\neq 0$},
\end{equation}
which implies, for any integer $\ell\in\mathbb{N}$, that
\begin{align}
&|\tilde{\eta}^+(t)|\leq \mathcal{C}_\ell |t|^{-\ell},\label{proeta+t}\\
&\mathcal{C}_\ell:=\frac{1}{2\pi}\int_{-\infty}^{+\infty}d\omega\,
\left|\frac{\partial^\ell}{\partial \omega^\ell}\eta^+(\omega)\right|. 
\end{align}

We consider a correlation function
\begin{align}
g(t):=\langle[\hat{A}(t),\hat{A}^\dagger]\rangle=\sum_{m,n}|\langle\Phi_m|\hat{A}|\Phi_n\rangle|^2\left(\rho(E_m)-\rho(E_n)\right)e^{i(E_m-E_n)t}.
\end{align}
On one hand, we have
\begin{align}
\int_{-\infty}^{+\infty} dt\, g(t)\tilde{\eta}^+(t)
&=\sum_{m,n}|\langle\Phi_m|\hat{A}|\Phi_n\rangle|^2
\left(\rho(E_m)-\rho(E_n)\right)\int_{-\infty}^{+\infty} dt\,e^{-i(E_n-E_m)t}\tilde{\eta}^+(t)\notag\\ 
&=\sum_{m,n}|\langle \Phi_m|\hat{A}|\Phi_n\rangle|^2\left(\rho(E_m)-\rho(E_n)\right)\eta^+(E_n-E_m)\notag\\
&\geq\sum_{m,n\,:\,2\varepsilon\leq E_n-E_m\leq K}|\langle \Phi_m|\hat{A}|\Phi_n\rangle|^2\left(\rho(E_m)-\rho(E_n)\right)\notag\\
&\geq\sum_{m,n\,:\,2\varepsilon\leq E_n-E_m\leq K}|\langle \Phi_m|\hat{A}|\Phi_n\rangle|^2\rho(E_m) \Delta.
\label{gintbound}
\end{align}
In passing to the third line, we used $\rho(E_n)< \rho(E_m)$ when $E_n> E_m$ and the conditions \eqref{etap} of $\eta^+(\omega)$.  In the last line, we defined
\begin{equation}
\Delta:=1-\max_{m,n\,:\,2\varepsilon\leq E_n-E_m \leq K}\frac{\rho(E_n)}{\rho(E_m)}.\label{defdelta}
\end{equation}
For the Gibbs state \eqref{Gibbs2}, we have
\begin{align}
&\Delta\geq h(\varepsilon)>0\quad\text{for $\varepsilon>0$},\label{deltabound}\\
&h(\varepsilon):=1-e^{-2\beta\varepsilon}.\label{hepsilon}
\end{align}

On the other hand, we can decompose the integral into two parts as  
\begin{equation}
\int_{-\infty}^{+\infty} dt\, g(t)\tilde{\eta}^+(t)
=\int_{|t|\ge T}dt\, g(t)\tilde{\eta}^+(t)+\int_{-T}^{+T}dt\, g(t)\tilde{\eta}^+(t)
\end{equation}
where $T$ is a large positive number.  \edd{For the first integral in the right-hand side}, we use the property Eq.~\eqref{proeta+t} of the function $\tilde{\eta}^+(t)$ as well as the trivial bound \edd{$|g(t)|\leq 2N_a^2$}~\footnote{Here and hereafter, we use the standard properties of the operator norm, such as
${|}{\langle}{\hat{O}}{\rangle}{|}\leq{\Vert}{\hat{O}}{\Vert}$, 
${\Vert}{\hat{O}}{\Vert}={\Vert}{\hat{O}}^\dagger{\Vert}$, and 
${\Vert}{\hat{O}\hat{O}'}{\Vert}\leq{\Vert}{\hat{O}}{\Vert}{\Vert}{\hat{O}'}{\Vert}$
for operators $\hat{O}$ and $\hat{O}'$}.  For a given function $\tilde{\eta}^+(t)$ with the parameters $\varepsilon$ and $K$, we can find a large $T$ such that  
\begin{equation}
\left|\int_{|t|\ge T}dt\, g(t)\tilde{\eta}^+(t)\right|\leq \varepsilon h(\varepsilon).\label{int2}
\end{equation}

\edd{For the second integral, we can use the Lieb-Robinson bound~\cite{LRB,HastingsReview}, from which we have~\cite{WO}
\begin{equation}
\Big\Vert \big[\hat{A}(t),\hat{A}^\dagger\big]\Big\Vert\leq \frac{C_1+C_2|t|^d}{|\Lambda|}
\end{equation}
for system-size-independent constants $C_1$ and $C_2$. Thus
\begin{align} 
\left|\int_{-T}^{+T}dt\, g(t)\tilde{\eta}^+(t)\right|
&\leq\int_{-T}^{+T}dt\, \big|\langle[\hat{A}(t),\hat{A}^\dagger]\rangle\big|\left|\tilde{\eta}^+(t)\right|\leq\int_{-T}^{+T}dt\, \frac{C_1+C_2|t|^d}{|\Lambda|}\frac{K}{2\pi}=K\frac{(d+1)C_1T+C_2T^{d+1}}{\pi(d+1)|\Lambda|},
\end{align}
where in the second inequality we used
\begin{align} 
\left|\tilde{\eta}^+(t)\right|\leq \frac{1}{2\pi}\int_{-\infty}^{+\infty}d\omega\, \eta^+(\omega)\leq\frac{1}{2\pi}\int_{\varepsilon}^{K+\varepsilon}d\omega\, 1=\frac{K}{2\pi}.
\end{align}
Therefore, for any given large $K$ and $T$, there exists a large volume $|\Lambda|$ such that }
\begin{align} 
\left|\int_{-T}^{+T}dt\, g(t)\tilde{\eta}^+(t)\right|\leq \varepsilon h(\varepsilon). \label{int1}
\end{align}

Combining Eqs.~\eqref{int2} and \eqref{int1} with the bound (\ref{gintbound}), we get
\begin{equation}
\label{difE+bound}
\sum_{m,n\,:\,2\varepsilon\leq E_n-E_m\leq K}
|\langle \Phi_m|\hat{A}|\Phi_n\rangle|^2\rho(E_m)\leq \frac{2\varepsilon h(\varepsilon)}{\Delta}\leq 2\varepsilon.
\end{equation}

\subsubsection{The range (ii): $-K\leq E_n-E_m\leq -2\varepsilon$}
Similarly, to estimate the contribution from the range $-K\leq E_n-E_m\leq -2\varepsilon$, we introduce a cutoff function $\eta^{-}\in C_0^\infty(\mathbb{R})$ that satisfies the following conditions:  
\begin{align}
\begin{cases}
\eta^{-}(\omega)=-1 &  (-K\leq\omega\leq -2\varepsilon),\\
\eta^{-}(\omega)=0 &  (\omega\leq -K-\varepsilon \text{ or }-\varepsilon\leq\omega),\\
-1\leq\eta^{-}(\omega)\leq0&  (\text{otherwise}).
\end{cases}
\end{align}
This time we have
\begin{align}
\int_{-\infty}^{+\infty} dt\, g(t)\tilde{\eta}^-(t)
&=\sum_{m,n}|\langle\Phi_m|\hat{A}|\Phi_n\rangle|^2
\left(\rho(E_m)-\rho(E_n)\right)\int_{-\infty}^{+\infty} dt\,e^{-i(E_n-E_m)t}\tilde{\eta}^-(t)\notag\\ 
&=\sum_{m,n}|\langle \Phi_m|\hat{A}|\Phi_n\rangle|^2\left(\rho(E_n)-\rho(E_m)\right)(-\eta^-(E_n-E_m))\notag\\
&\geq\sum_{m,n\,:\,-K\leq E_n-E_m\leq -2\varepsilon}|\langle \Phi_m|\hat{A}|\Phi_n\rangle|^2\left(\rho(E_n)-\rho(E_m)\right)\notag\\
&\geq\sum_{m,n\,:\,-K\leq E_n-E_m\leq -2\varepsilon}|\langle \Phi_m|\hat{A}|\Phi_n\rangle|^2\rho(E_n) \Delta\notag\\
&\geq\sum_{m,n\,:\,-K\leq E_n-E_m\leq -2\varepsilon}|\langle \Phi_m|\hat{A}|\Phi_n\rangle|^2\rho(E_m) \Delta.
\end{align}
Here, $\Delta$ is defined in Eq.~\eqref{defdelta}. In the same way as before, we find
\begin{equation}
\label{difE-bound}
\sum_{m,n\,:\,-K\leq E_n-E_m\leq -2\varepsilon}
|\langle \Phi_m|\hat{A}|\Phi_n\rangle|^2\rho(E_m)\leq  2\varepsilon.
\end{equation}

\subsubsection{The range (iii): $K<|E_n-E_m|$}
The third contribution can be easily bounded by using a trick.
\begin{align}
\sum_{m,n\,:\,K<|E_n-E_m|}|\langle\Phi_m|\hat{A}|\Phi_n\rangle|^2\rho(E_m)
&\leq \frac{1}{K^2}\sum_{m,n}(E_n-E_m)^2|\langle\Phi_m|\hat{A}|\Phi_n\rangle|^2\rho(E_m)= \frac{1}{K^2} \langle [\hat{A},\hat{H}][\hat{H},\hat{A}^\dagger]\rangle\notag\\
&\leq \edd{\frac{1}{K^2} \big\Vert [\hat{A},\hat{H}]\big\Vert^2.}
\end{align}
Thanks to the assumed locality of the Hamiltonian, \edd{this operator norm can be bounded as
\begin{align}
&\big\Vert [\hat{A},\hat{H}]\big\Vert=\frac{1}{|\Lambda|}\sum_{\vec{x},\vec{y}\in\Lambda\,:\,|\vec{x}-\vec{y}|\leq R_h+R_a}\big\Vert [\hat{a}_{\vec{x}},\hat{h}_{\vec{y}}]\big\Vert\leq C,\\
&C:= 2N_aN_hv(R_a+R_h),\\
&v(R):=\frac{1}{|\Lambda|}\sum_{\vec{x},\vec{y}\in\Lambda\,:\,|\vec{x}-\vec{y}|\leq R}1.\label{vr}
\end{align}
Note that $v(R)$ does not grow with $|\Lambda|$.}  Therefore,
\begin{equation}
\label{difElargeK}
\sum_{m,n\,:\,K<|E_n-E_m|}|\langle\Phi_m|\hat{A}|\Phi_n\rangle|^2\rho(E_m)\leq\frac{C}{K^2}.
\end{equation}

\subsubsection{The range (iv): $|E_n-E_m|<\varepsilon$}
Finally, using the fact that $|e^{ix}-1|=2|\sin \frac{x}{2}|\leq |x|$ for any real number $x$, we get
\begin{align}
\sum_{m,n\,:\,|E_n-E_m|<2\varepsilon}|\langle \Phi_m|\hat{A}|\Phi_n\rangle|^2\rho(E_m)\left|e^{i(E_m-E_n)t}-1\right|
&\leq 2\varepsilon|t|\sum_{m,n:\,|E_n-E_m|<2\varepsilon}|\langle \Phi_m|\hat{A}|\Phi_n\rangle|^2\rho(E_m)\notag\\
&\leq 2\varepsilon|t|\sum_{m,n}|\langle \Phi_m|\hat{A}|\Phi_n\rangle|^2\rho(E_m)=2\varepsilon|t| \langle \hat{A}\hat{A}^\dagger\rangle\notag\\
&\edd{\leq 2\Vert \hat{A}\Vert^2\varepsilon|t|  \leq 2N_a^2\varepsilon |t|.}\label{smallEdef}
\end{align}
%
This completes the verification of Eq.~\eqref{combine} and hence the proof of Eq.~\eqref{result2}.

\subsection{Proof for $\beta=0$}
\label{secbeta0}
Interestingly, the proof in the previous section does not apply to the infinite temperature ($\beta=0$) where \ed{$\rho(E_n)$ in Eq.~\eqref{Gibbs2} becomes constant:
\begin{equation}
\rho(E_n)=\frac{1}{\mathcal{D}}.
\end{equation}
Here $\mathcal{D}$ is the dimension of the entire Hilbert space. }
In this special case, however, we can directly prove Eq.~\eqref{result}
using the \ed{clustering property of the infinite-temperature state~\footnote{In the earlier version of the manuscript~\cite{v1}, our proof for $\beta=0$ was based on the Lieb-Robinson bound. The present simpler proof was informed by Yichen Huang. \edd{See also Ref.~\cite{Huang} and Sec.~\ref{sec.Discussion}}}.} 
Thus the ``absence of the time crystals'' also holds at the
infinite temperature, consistently with the intuition that
the infinite temperature is the most disordered limit.


At $\beta=0$, the equal-time correlation function trivially exhibits
the locality
\begin{equation}
 \langle \hat{a}_{\vec{x}} \hat{b}_{\vec{y}} \rangle
 =  
 \langle \hat{a}_{\vec{x}} \rangle
 \langle \hat{b}_{\vec{y}} \rangle
\label{eq.locality}
\end{equation}
if the support of $\hat{a}_{\vec{x}}$ and $\hat{b}_{\vec{y}}$ do not overlap.  This implies the clustering property
and the absence of any spatial long-range order.
However, quantum dynamics is nontrivial even at the
infinite temperature (see, for example, Ref.~\cite{Znidaric-PRL2011}
and references therein)
and the question of the time crystal is not totally trivial.

\ed{To prove Eq.~\eqref{result2} in this setting, let us define
\begin{align}
\delta\hat{A}:=\hat{A}-\langle \hat{A}\rangle=\frac{1}{|\Lambda|}\sum_{\vec{x}\in\Lambda} (\hat{a}_{\vec{x}}-\langle \hat{a}_{\vec{x}}\rangle).
\end{align}
It follows that
\begin{align}
\Big|\langle\delta\hat{A}(t)\,\delta\hat{A}^\dagger\rangle\Big|=\Big|\sum_{n,m}|\langle \Phi_n|\delta\hat{A}|\Phi_m\rangle|^2 \rho(E_n)e^{i(E_n-E_m)t}\Big|\leq \sum_{n,m}|\langle \Phi_n|\delta\hat{A}|\Phi_m\rangle|^2\rho(E_n)=\langle\delta\hat{A}\,\delta\hat{A}^\dagger\rangle\label{Schwartz}
\end{align}
and that
\begin{align}
\Big|\langle\hat{A}(t)\hat{A}^\dagger\rangle-\langle\hat{A}\,\hat{A}^\dagger\rangle\Big|=\Big|\langle\delta\hat{A}(t)\,\delta\hat{A}^\dagger\rangle-\langle\delta\hat{A}\,\delta\hat{A}^\dagger\rangle\Big|\leq 2 \langle\delta\hat{A}\,\delta\hat{A}^\dagger\rangle.\label{Schwartz2}
\end{align}
The locality in Eq.~\eqref{eq.locality} implies that $\langle\delta\hat{A}\,\delta\hat{A}^\dagger\rangle$ is inversely proportional to $|\Lambda|$. Therefore, we obtain Eq.~\eqref{result2} which gives Eq.~\eqref{result} as explained in Sec.~\ref{outline}.}

\section{Discussion}
\label{sec.Discussion}
The proof for the Gibbs state \ed{at a finite temperature in Sec.~\ref{secbetap}} equally applies to other stationary states \ed{($[\hat{\rho},\hat{H}]=0$)} of a local, static Hamiltonian
as long as the density operator $\hat{\rho}$ satisfies the following conditions.
\begin{enumerate}
\item 
The weight $\rho(E_n)$ for each eigenstate $|\Phi_n\rangle$ should be a strictly decreasing function of $E_n$, i.e.,  $\rho(E_n)<\rho(E_m)$ when $E_n>E_m$ for any $n$ and $m$.
\item There exists a smooth function $h(\varepsilon)$ of $\varepsilon$ such that the quantity $\Delta$ defined in Eq.~\eqref{defdelta} is bounded below as in Eq.~\eqref{deltabound}. Furthermore, $h(\varepsilon)$ must be independent of the system size $|\Lambda|$. 
\end{enumerate}
\ed{This argument can also} be straightforwardly modified when
the weight $\rho(E_n)$ is a strictly increasing function of $E_n$.
Therefore, the same theorem holds for Gibbs state with
a ``negative temperature'' ($\beta <0$), which is well-defined
for a bounded Hamiltonian we discuss here.


\ed{Recently, \edd{in the interesting paper~\cite{Huang},
Huang showed the inequality}
\begin{align}
\left|\langle\hat{A}(t)\hat{B}\rangle-\langle\hat{A}\hat{B}\rangle\right|=
O(|\Lambda|^{-1})
\end{align}
for \edd{an arbitrary} stationary state $\hat{\rho}$ assuming a sufficiently fast decay of \edd{spatial} correlation functions but without assuming the locality of the Hamiltonian.
Here we \edd{comment} that, 
\edd{ in order to prove the weaker statement Eq.~\eqref{result},
which already implies the absence of long-range temporal order,
it is sufficient to assume the ``clustering'' property of
spatial correlation functions.}
We say a state exhibits \emph{clustering} if there exists a system-size independent function $f(r)$ with $\lim_{r\rightarrow\infty}f(r)=0$ such that correlation functions of any local bounded operator obey
\begin{align}
|\langle \delta \hat{a}_{\vec{x}}\,\delta\hat{a}_{\vec{y}}^\dagger\rangle|\leq f(|\vec{x}-\vec{y}|)\quad\text{for all}\quad\vec{x}, \vec{y}\in\Lambda.\label{clustering}
\end{align}
As we discuss in Appendix~\ref{app}, Eq.~\eqref{clustering} readily gives
\begin{align}
\lim_{|\Lambda|\rightarrow\infty}\langle\delta\hat{A}\,\delta\hat{A}^\dagger\rangle=0.\label{statement2}
\end{align}
Moreover, Eq.~\eqref{Schwartz} holds as long as $\rho(E_n)\geq0$ for all $n$ even when $\rho(E_n)$ is not constant~\cite{Huang}.  Thus, we get Eq.~\eqref{result2} by combining Eqs.~\eqref{Schwartz2} and \eqref{statement2}. }

\edd{In conclusion, in order to realize a temporal long-range order in a stationary state, the system has to fulfill both of the following two conditions:
\begin{enumerate}
\item Either the weight $\rho(E_n)$ breaks some of the above conditions or the Hamiltonian $\hat{H}$ violates the locality.
\item The state $\hat{\rho}$ does not possess the clustering property.
\end{enumerate}
For example, a time-crystalline behavior may be observed in a single eigenstate (except for the ground state) or a micro-canonical ensemble of a local Hamiltonian (e.g. Ref.~\cite{Syrwid-TC-Excited2017}), and in the ground state of a non-local Hamiltonian (e.g. Ref.~\cite{PhysRevLett.123.210602}).}

\vspace{1.0\baselineskip}
{\bf Note added in proof:} After we revised our manuscript and updated the version on arXiv, Huang also posted the second version of his paper (Y. Huang, arXiv:1912.01210v2) in which the author made a revision for the main result in the first version~\cite{Huang}, and added a new result which is similar to ours in Sec. III.

\begin{acknowledgments}
We thank Hal Tasaki for fruitful discussions. 
We also thank Shivaji Sondhi for encouraging us to publish the present result, which is based on an earlier unpublished note and \ed{Yichen Huang for informing us of a simplification of the proof for $\beta=0$.}
The work of H.~W. was supported by JSPS KAKENHI Grant No.~JP17K17678 and by JST PRESTO Grant No.~JPMJPR18LA.
The work of M.~O. was supported in part by
JSPS KAKENHI Grant No.~JP19H01808 and by
US National Science Foundation Grant No.~NSF PHY-1748958
through Kavli Institute for Theoretical Physics, UC Santa Barbara.
\end{acknowledgments}

\appendix
\section{Fluctuation and clustering}
\label{app}
In this appendix, we show that fluctuations of normalized macroscopic observables are negligible in the large volume limit when the state possesses the clustering property.  The clustering defined in Eq.~\eqref{clustering} means that, for any $\varepsilon>0$, there exists $R>0$ (independent of $|\Lambda|$) such that 
\begin{align}
|\langle \delta \hat{a}_{\vec{x}}\,\delta\hat{a}_{\vec{y}}^\dagger\rangle|<\frac{1}{2}\varepsilon\quad\text{if}\quad |\vec{x}-\vec{y}|>R.
\end{align}
It follows that  
\begin{align}
\langle\delta\hat{A}\,\delta\hat{A}^\dagger\rangle&\leq\frac{1}{|\Lambda|^2}\sum_{\vec{x},\vec{y}\in\Lambda\,:\,|\vec{x}-\vec{y}|\leq R}| \langle \delta \hat{a}_{\vec{x}}\,\delta\hat{a}_{\vec{y}}^\dagger\rangle|+\frac{1}{|\Lambda|^2}\sum_{\vec{x},\vec{y}\in\Lambda\,:\,|\vec{x}-\vec{y}|> R}| \langle \delta \hat{a}_{\vec{x}}\,\delta\hat{a}_{\vec{y}}^\dagger\rangle|\leq \frac{2N_a^2v(R)}{|\Lambda|}+\frac{1}{2}\varepsilon,
\end{align}
where $v(R)$ is defined in Eq.~\eqref{vr}.  Since $N_a$ is independent of the system size, we can find $|\Lambda|$ such that $\frac{2N_a^2v(R)}{|\Lambda|}<\frac{1}{2}\varepsilon$. Therefore, for a sufficiently large system size, we have
\begin{align}
\langle\delta\hat{A}\,\delta\hat{A}^\dagger\rangle<\varepsilon.
\end{align}
This completes the proof of Eq.~\eqref{statement2}.

\end{document}